\documentclass[12pt,preprint]{aastex}

\slugcomment{ApJ accepted}

\shorttitle{\ion{Fe}{2} Diagnostic Tools}
\shortauthors{Verner et al.}

\begin{document}


\title{\ion{Fe}{2} Diagnostic Tools for Quasars}


\author{E. Verner\altaffilmark{1,2,3}
and F. Bruhweiler\altaffilmark{1,2,4}}
\altaffiltext{1}{IACS/Dept. of Physics, Catholic University of America.}
\altaffiltext{2}{Laboratory of Astronomy and Solar Physics, NASA/Goddard Space 
Flight Center Greenbelt MD 20771.}
\altaffiltext{3}{email: kverner@fe2.gsfc.nasa.gov}
\altaffiltext{4}{email: fredb@iacs.gsfc.nasa.gov}

\author{D. Verner\altaffilmark{1,5}}
\altaffiltext{5}{email: verner15@comcast.net}

\author{S. Johansson\altaffilmark{6}}
\altaffiltext{6}{Lund Observatory, Lund University, P.O. Box 43, S-22100 
Lund, Sweden, Sveneric.Johansson@astro.lu.se}

\author{T. Kallman \altaffilmark{7}}
\altaffiltext{7}{Laboratory for High Energy Astrophysics, NASA Goddard Space
Flight Center, Greenbelt, MD20771,
timothy.r.kallman@nasa.gov}

\author{T. Gull \altaffilmark{2,8}}
\altaffiltext{8}{email: theodore.r.gull@nasa.gov}


\begin{abstract}

The enrichment of Fe, relative to alpha-elements such as O and Mg, represents a
potential means to determine the age of
quasars and probe the galaxy formation epoch.

To explore how \ion{Fe}{2} emission in quasars is linked to physical
conditions and abundance, we have constructed a 830-level \ion{Fe}{2} model
atom and investigated through photoionization calculations
how \ion{Fe}{2} emission  strengths depend on
non-abundance factors.
We have split \ion{Fe}{2} emission into three major wavelength bands,
\ion{Fe}{2} (UV), \ion{Fe}{2}(Opt1),
and \ion{Fe}{2}(Opt2),
and explore how the  \ion{Fe}{2}(UV)/\ion{Mg}{2},
\ion{Fe}{2}(UV)/\ion{Fe}{2}~(Opt1)
and  \ion{Fe}{2}(UV)/\ion{Fe}{2}~(Opt2) emission ratios depend upon
hydrogen density and ionizing flux in broad-line regions (BLR's) of 
quasars.  Our
calculations show that:
1) similar \ion{Fe}{2}(UV)/\ion{Mg}{2} ratios can exist over a wide
range of  physical conditions;
2) the \ion{Fe}{2}(UV)/\ion{Fe}{2}~(Opt1) and
\ion{Fe}{2}(UV)/\ion{Fe}{2}~(Opt2) ratios serve to constrain ionizing
luminosity and hydrogen density; and
3)  flux measurements of \ion{Fe}{2} bands and knowledge of ionizing flux provide
tools to derive distances to BLR's in quasars.
To derive all BLR physical parameters with uncertainties, 
comparisons of our model with observations of a large quasar 
sample at low redshift ($z<1$) is desirable.

The  STIS and NICMOS spectrographs aboard the Hubble Space Telescope 
(HST) offer the best means to provide such observations.

\end{abstract}




\section{Introduction}
The observed ratio of restframe \ion{Fe}{2}~UV emission from 
2200$-$3000~{\AA} 
(thereafter, \ion{Fe}{2}(UV)) to 
that of the \ion{Mg}{2} 2800~{\AA} resonance doublet has been widely used to 
estimate Fe/Mg abundance ratios in BLRs of quasars (Wills et al. 1985, 
thereafter WNW85; Iwamuro et al.  2002; Dietrich et al. 2002, 2003;
Freudling et al. 2003; Barth et al. 2003;
 Maiolino et al. 2003). 
Iron enrichment is not expected in evolution scenarios at 
redshifts $z \leq 1$ (Hamann \& Ferland 1993; Yoshii et al. 1996; 
Heger \& Woosley 2002). Meanwhile, there is some  
observational evidence that narrow line AGN have higher 
metalicities compared to other low redshift AGN (Shemmer \& Netzer 2002).

Measurements of
\ion{Fe}{2}(UV)/\ion{Mg}{2}  emission ratios in quasars show large scatter 
from 1 to 20, and no redshift dependency up to $z \sim 6.4$ (Iwamuro et al. 2002).

In a recent paper, we have used a sophisticated 830-level model for 
\ion{Fe}{2}   to investigate abundance and microturbulence effects
on these ratios (Verner et al. 2003a). Our modeling indicates strong \ion{Fe}{2} 
emission from 1000 to 6800~\AA, and reveals that
\ion{Fe}{2} (UV)/\ion{Mg}{2} ratios (at {$n_H$ }= 10$^{9.5}$~cm$^{-3}$ 
and total
hydrogen column
density $N_{H}$ = 10$^{24 }~$cm$^{-2}$) increase from 5 to 30 for
microturbulence $v_{turb}$ varied from 1 to 100 km~s$^{-1}$, 
while an increase in abundance
by factor 10 only increases the same ratios by factor 2. 
We have further found that a reasonable range of microturbulence is 
between 5$-$10 km ~s$^{-1}$,
and \ion{Fe}{2} optical emission flux from 4000$-$6000~{\AA}  is more sensitive to 
abundance than is the \ion{Fe}{2}(UV) band.
Conversely, the  \ion{Fe}{2}(UV) band is more sensitive to  microturbulence than the
\ion{Fe}{2} optical band.

In this paper we continue our study of \ion{Fe}{2} emission line
formation in BLRs of quasars in an attempt to explain the nature
of the large observed scatter. Since abundances, density,  
and excitation conditions are ill-determined in BLRs,
we have performed a set of numerical calculations to study the 
effects of varying hydrogen and 
ionizing photon densities. 
We have split \ion{Fe}{2} emission into three  wavelength bands 
that are most commonly 
measured in observations:
\ion{Fe}{2}(UV) (2000$-$3000~{\AA}),
\ion{Fe}{2}(Opt1) (3000$-$3500~{\AA}),
and \ion{Fe}{2}(Opt2) (4000$-$6000~{\AA}).
The major goal is to ascertain if \ion{Fe}{2}(UV)/\ion{Mg}{2},
\ion{Fe}{2}(UV)/\ion{Fe}{2}(Opt1),
and \ion{Fe}{2}(UV)/\ion{Fe}{2}(Opt2) emission ratios present different
trends due to non-abundance factors.

Although many factors contribute to increase \ion{Fe}{2} emission fluxes, 
the effects can be separated by using several ratios simultaneously.
Spectra of quasars at redshifts up to $z = 1$ are especially valuable, 
since they should exhibit no elemental overabundances, and the effects 
of physical parameters such as hydrogen density, ionizing luminosity, 
and even microturburbulence can be easily explored. We just showed 
general trends, while a detailed comparison between model and 
observations for each quasar spectrum should be done individually.
The figures in our paper represent calculations for a range of 
reasonable BLR conditions, but do not cover the full range of 
possible variables and should be used with extreme care by observers. 
Such comparison will provide the first fundamental test of the \ion{Fe}{2} 
emission model. This consistency check is a necessary step 
before one can confidently apply this methodology to quasars 
over a wide range of redshift as a means to probe galactic 
chemical evolution models.
The required data for quasars up to $z = 1$ can be obtained 
with STIS and NICMOS aboard the HST.

\section{Non-abundance Effects and BLR Diagnostic Based on 
\ion{Fe}{2} Emission}
Why quasars at moderate to high redshift exhibit strong UV \ion{Fe}{2} 
emission is one the unsolved problems of AGN studies. 
The extremely complex energy level structure (Johansson 1978) of \ion{Fe}{2} 
makes it very difficult to obtain all the experimental transition probabilities 
and, therefore, calculate line intensities. Several
hundreds of transitions of \ion{Fe}{2} must be considered, many with
large optical depths. Our earlier model for Fe$^{+}$ included 371 energy levels 
below 11.6 eV (Verner et al. 1999),which  was incorporated into photoionization 
code CLOUDY (Ferland et al. 1998). Even though the upper energy levels 
in the current model (Verner et al. 2003a) have been extended only
about 2.5 eV higher compared to the old model,
the total number of transitions has increased dramatically from 68,638 to
344,035. The model includes 830 levels up to 14.1 eV. 
This increase in transitions is mainly due to increased density of energy levels 
at higher energy. The energy level data are  from Johansson (2004).

The large number of  \ion{Fe}{2} lines form several distinct emission bands 
recognized in  early observational work (e.g. Greenstein \& Schmidt 1964;
Wampler \& Oke 1967; Sargent 1968; Netzer \& Wills 1983)
and theoretical modeling (WNW85; Verner et al. 1999).
The 830-level ion model is far more accurate than the previous best efforts.
The increased number of Fe$^+$ energy levels influences not only the
\ion{Fe}{2} spectrum but also the whole energy budget, 
temperature, and consequently line emission of other elements in the emitting region.
However, an in-depth study of the new theoretical emission spectrum will 
not be given in this paper.

For several reasons, we have followed a more general approach
and considered three wide \ion{Fe}{2} bands in our
calculations, namely \ion{Fe}{2} (UV),
\ion{Fe}{2}~(Opt1), and \ion{Fe}{2}~(Opt2). The biggest unknown factor in
predicting \ion{Fe}{2} (UV) and
\ion{Mg}{2}  emission is the magnitude
of the velocity of turbulence in BLRs. Whether it is
on the order of 10 ~km~s$^{-1}$, 100 ~km~s$^{-1}$ or
even higher is not at all clear (Alexander \& Netzer 1997; Murray \& Chiang
1997).
Also, the \ion{Fe}{2} emission in BLRs is present over a wide wavelength 
range from 1000~{\AA} to the IR. Because of the velocity broadening, 
presumably due to orbital motion, the \ion{Fe}{2} emission can be characterized 
by a pseudo-continuum superposed upon the intrinsic power-law spectrum of 
the quasar.

For the adopted parameter range, we have investigated how hydrogen 
density
and photon density of hydrogen ionizing photons
at the illuminated face alter the intrinsic emission ratios of
\ion{Fe}{2}(UV)/\ion{Mg}{2},  \ion{Fe}{2} (UV)/\ion{Fe}{2}(Opt1)
and \ion{Fe}{2} (UV)/\ion{Fe}{2}(Opt2).
For our calculations, we have used the same \ion{Fe}{2} energy level structure
and model as in Verner et al. (2003a).
We have looked for the variations of \ion{Fe}{2}(UV)/\ion{Mg}{2} emission
ratios  in BLRs assuming solar abundance
for a wide range of hydrogen density, {$n_H$ }= 10 $^{9.5} - 10^{13.0}$
cm$^{-3}$, and total column
density, $N_{H}$ = 10$^{24 }~$cm$^{-2}$. We further assume that the flux of
hydrogen ionizing photons
at the illuminated face is 10$^{17.5} - 10^{22.0}~$cm$^{-2}$
s$^{-1}$.  These parameters for BLR conditions
are within the range of values taken from Verner et al. (1999) and Verner
(2000). We employ the characteristic AGN
continuum described in Korista et al. (1997), which consists of a UV bump
peaking
near 44 eV, a $f_{\nu} \propto \nu^{-1}$ X-ray power law, and a UV to
optical spectral index, $a_{ox}=  -1.4$.

Our knowledge about turbulence is very limited.
Consequently, in our initial calculations of \ion{Fe}{2} emission, we have 
produced models with microturbulence with velocities of
$v_{turb}~=~0,~5, 10,$ and 100 ~km~s$^{-1}$.
We find that the strength of the \ion{Fe}{2}(UV) emission is very 
sensitive 
to microturbulence. Non-zero turbulence velocities
help to explain the observed smooth shape of \ion{Fe}{2}(UV), and
a $v_{turb}~=~5$~km~s$^{-1}$ makes the 
model fit reasonable (Verner et al. 2003b)
 \ion{Fe}{2} emission strengths
forming a pseudo-continuum (Figure 1) at $v_{turb}~=~5$~km~s$^{-1}$.
Even in emission lines, this must be a curve-of-growth effect. Stronger 
lines (higher A-values) should have larger equivalent width and be more 
sensitive to microturbulence than weaker lines.

Figure 2 shows plots of \ion{Fe}{2}(UV)/\ion{Mg}{2} ratios versus
hydrogen density and flux of ionizing photons 
at $v_{turb}~=~0,~5, $ and 10 ~km~s$^{-1}$. 

We see from Figure 2 that at low flux, $ \Phi < 10^{19.0}$~cm$^{-2}$~s$^{-1}$, 
the \ion{Fe}{2}(UV)/\ion{Mg}{2} $\sim 1$ and 
it does not depend on turbulent velocity. The increase of 
microturbulence works similar to a hydrogen density increase, and large 
\ion{Fe}{2}(UV)/\ion{Mg}{2} ratios are predicted at smaller densities.

Although solar abundance is assumed 
throughout,  the possible range of values of \ion{Fe}{2}(UV)/\ion{Mg}{2} are quite large, 
from 1 to 40 (Fig. 2, $v_{turb}~=~5$~km~s$^{-1}$). Fig. 2 also demonstrates that the same
\ion{Fe}{2}(UV)/\ion{Mg}{2} ratios may indicate a wide range of 
physical conditions. At densities below {$n_H$ }= 10$^{11.0}$ cm$^{-3}$,
the dependence on luminosity displays a different trend compared to that at larger 
densities.
The \ion{Fe}{2}(UV)/\ion{Mg}{2} ratio reaches a maximum near {$n_H$ }= 10
$^{10.0}$ cm$^{-3}$.
In quasars, if hydrogen density is less than {$n_H$} = 10$^{11.0}$ cm$^{-3}$, 
all \ion{Fe}{2}(UV)/\ion{Mg}{2} values are less than 8. 
Small \ion{Fe}{2}(UV)/\ion{Mg}{2} ratios (up to 2) are possible indicators of two 
different regimes: a) high luminosity conditions at small densities 
({$n_H$ } $\leq$ 10 $^{11.0}$ cm$^{-3}$) or b) low-luminosity conditions over  
a wide density range.
In the latter case, these ratios are insensitive to density variations.
The left upper corner in Figure 2 corresponds to low hydrogen density and large 
ionizing flux, and shows that iron and magnesium are effectively more highly ionized, beyond \ion{Fe}{2} and \ion{Mg}{2}. The large ionizing flux at 
high hydrogen density is insufficient to ionize \ion{Fe}{2}. Instead, this 
combination increases the strength of the \ion{Fe}{2}(UV) emission 
band relative to the  
\ion{Mg}{2}  doublet emission. As a result, large \ion{Fe}{2}(UV)/\ion{Mg}{2} 
ratios (from 8 to 40) are predicted for large hydrogen density 
{$n_H = 10^{11.0} - 10^{13.0}$ cm$^{-3}$ and ionizing fluxes 
$10^{20.5} - 10^{22.0}~$cm$^{-2}$ s$^{-1}$. If the scatter observed at any given 
redshift is due to variations in density and luminosity, it may well mask any 
abundance effect.

As we have already shown (Verner et al. 2003a), the
\ion{Fe}{2}(UV)/\ion{Fe}{2}(Optical) ratio is less sensitive to microturbulence 
velocity than is
the \ion{Fe}{2}(UV)/\ion{Mg}{2}  ratio. 
Figures 3 and 4 illustrate this conclusion for both
ratios, namely \ion{Fe}{2}(UV)/\ion{Fe}{2}(Opt1) and \ion{Fe}{2}(UV)/\ion{Fe}{2}(Opt2).

Figure 3 shows how the \ion{Fe}{2}(UV) and \ion{Fe}{2}(Opt1) emission varies versus 
physical conditions at $v_{turb}~=~0,~5, $ and 10 ~km~s$^{-1}$. 
If $v_{turb}~=~5 $~km~s$^{-1}$, the \ion{Fe}{2}(UV)/\ion{Fe}{2}(Opt1) ratios vary from $\sim 1$ 
to 30.
However, there is a wide plateau with almost constant \ion{Fe}{2}(UV)/\ion{Fe}{2}(Opt1) 
ratio $6 - 7$. At low ionizing flux with increasing density, \ion{Fe}{2}(UV) dominates 
over \ion{Fe}{2}(Opt1). Ratios $\geq 8$ show a dependence on physical conditions and 
are more suitable to use as a diagnostic of ionizing photon flux and hydrogen number 
density.

The \ion{Fe}{2}(UV)/\ion{Fe}{2}(Opt2) ratios exhibit a much stronger 
dependence on density than either \ion{Fe}{2}(UV)/\ion{Mg}{2} 
or \ion{Fe}{2}(UV)/\ion{Fe}{2}(Opt1).
The plateau where ratios are insensitive to physical conditions is much smaller
than that in the previous case. Similarly, the 
\ion{Fe}{2}(UV) dominates over \ion{Fe}{2}(Opt2) emission strength in the domain 
of high density and low ionizing flux (Fig. 4 at $v_{turb}~=~5 $ ~km~s$^{-1}$).

The \ion{Fe}{2} bands include different \ion{Fe}{2} transitions.
The \ion{Fe}{2}(UV) band includes the strongest UV multiplets  from UV1 to UV5. 
The \ion{Fe}{2}(Opt1) band includes optical multiplets 1, 4-7 and \ion{Fe}{2}(Opt2)
optical multiplets 21, 22, 25, 27, 28, 35-38. UV multiplets are due to transitions 
from higher energy upper levels compared to those of the 
optical multiplets. Therefore, UV multiplets become stronger with 
increasing density 
due to increase of upper level populations,
and \ion{Fe}{2}(UV)/\ion{Fe}{2}(Opt1), \ion{Fe}{2}(UV)/\ion{Fe}{2}(Opt2) ratios 
become larger. 

While no single ratio can constrain physical conditions, their combination can provide 
valuable 
diagnostics of density and ionizing flux. If our assumption about microturbulence is correct, 
we might well observe 
large \ion{Fe}{2}(UV)/\ion{Mg}{2} values ($>10$) at solar abundance.
For the same physical conditions, \ion{Fe}{2}(UV)/\ion{Fe}{2}(Opt1) ratios will be in 
a range of $5 - 10$ and \ion{Fe}{2}(UV)/\ion{Fe}{2}(Opt2) in a range of $7 - 15$. Likewise, large scatter in \ion{Fe}{2}(UV)/\ion{Mg}{2}, from 1 to 30, can be explained by non-abundance factors, where both ionizing luminosity and hydrogen density are responsible. 
Values of \ion{Fe}{2}(UV)/\ion{Mg}{2} $\geq$ 10  are most 
probably achieved at hydrogen densities, larger than {$n_H$ } = 10 $^{11.0}$ cm$^{-3}$ 
at high ionizing flux.

It is highly desirable to develop an observational template to derive Mg/Fe abundances. Yet, it is not clear that any single template can be constructed that would be appropriate for all quasar spectra. A combined observational and detailed modeling program is needed to determine the answer to this question. Currently, two types of \ion{Fe}{2} templates have been used. One template is based upon the I~ZW~1 spectrum (e.g. Vestergaard \& Wilkes 2001; Dietrich et al. 2002, 2003) and the other  is based on an average quasar template (WNW85, Iwamuro et al. 2002).
Our comparisons between model and observations reveal that the I~ZW~1 template
by Vestergaard \& Wilkes (2001) is not generally applicable to BLRs due to variable \ion{Fe}{2} emission   contributions contaminating measurements of the \ion{Mg}{2} doublet.   
The contributions of Fe II lines are easily seen in our modeling of BLRs. Failure to account for this \ion{Fe}{2} contamination to the \ion{Mg}{2} emission can lead to erroneous \ion{Mg}{2} emission estimates. 
Assuming $\log \Phi_{ion} = 20.5$, the values of \ion{Fe}{2} 
(2720 $-$ 2860~{\AA})/ \ion{Mg}{2} are 1.3, 0.8 and 0.2, 
for $n_H = 10^{11}$ cm$^{-3}$, $10^{12}$ cm$^{-3}$, and 
$10^{13}$ cm$^{-3}$, respectively. Thus, errors in the 
\ion{Mg}{2} flux can be of factor 2 and even higher, if the 
contribution of \ion{Fe}{2} lines is not accounted for. Naturally, 
the larger measured \ion{Fe}{2} (UV)/ \ion{Mg}{2} ratio the more 
\ion{Mg}{2} affected by the \ion{Fe}{2} emission (Fig. 5).

The emission spectrum of \ion{Fe}{2}  in BLRs  spans a wide wavelength range from the UV to IR. It affects not only line intensity measurements of other ions, 
but the continuum determinations as well. Based on the model predictions, it is possible to define a narrow ($\sim 20$~{\AA}) \ion{Fe}{2} window in quasar spectra near 3,050~{\AA}, where little or no \ion{Fe}{2} emission is present.
In case of the strong \ion{Fe}{2}(UV) band, the emission produces a pseudo-continuum that extends  to wavelengths as short as 1,000~{\AA}.
How much the \ion{Fe}{2}(UV) pseudo-continuum affects the 1,000~{\AA} $-$
2,000~{\AA} range, depends on microturbulence and hydrogen density,  and is the  subject of future investigation. If any of these \ion{Fe}{2} windows are used to determine the underlying quasar continuum, it must be done with caution.




In most studies of quasars, even at low redshift, Fe is thought to have super-solar abundances.
WNW85 were the first who recognized that ultraviolet \ion{Fe}{2} emission contributed to \ion{Mg}{2} resonance doublet.
However, in their analysis they still needed an iron overabundance of approximately 3 times solar to explain the high \ion{Fe}{2}(UV)/\ion{Mg}{2} ratios.
  From comparisons of Figures $2-4$ we can deduce the physical conditions in
BLRs. 
The ionizing flux at the face of the cloud is $\Phi_{ion} = L_{ion}/4\pi r^2$, where 
$L_{ion}$ is total ionizing luminosity.
Since we can easily estimate the ionizing luminosity from observations, we can calculate the number 
of ionizing photons at any distance from the central
source, and hence obtain a distance for the BLR from the central engine.

We selected five objects from WNW85 with measured \ion{Mg}{2} and 
\ion{Fe}{2}(UV), \ion{Fe}{2}(Opt1), and \ion{Fe}{2}(Opt2) emission bands.
We then compared these measurements with Figures 2$-$4 to 
deduce hydrogen densities and ionizing photon fluxes for the BLRs of the quasars. Finally, 
we have ascertained if enhanced Fe abundances are required to explain the observations. 
Where possible, we have used the values corrected for intrinsic and foreground reddening. 
The results are given in Table 1. Column 1 shows the name of the object, where the three following columns present the observed emission ratios similar to that calculated in our model. The next two columns list hydrogen densities and ionizing photon fluxes. 
The last two columns give photon luminosities and derived sizes of emitting regions.

When we compare these results with the WNW85 models, we find no iron overabundance at redshifts $0.15 < z < 0.7$. The best fits are obtained within reasonable hydrogen densities, $10^{10.0} - 10^{13.0}$ cm$^{-3}$. 
The much larger number of \ion{Fe}{2} levels used in the photoionization calculations provides more sensitivity to the radiation field, and  enables us to explain the much larger observed \ion{Fe}{2}(UV)/\ion{Mg}{2} ratios.

Quasar luminosities were derived from energy fluxes measured by the ASCA X-ray 
observations (the Tartarus  database; Turner \& Nandra, http://tartarus.gsfc.nasa.gov/) for 
Q0405$-$123 and Q1226+023, and from the Rosat radio loud catalog 
(Brinkmann et al. 1997) for the remainder of the objects.  
These were converted into 0.0136$-$10 keV number fluxes using the power law indices, 
and then into number photon luminosity assuming  {H$_0$}~=~71 km~s$^{-1}$~Mpc$^{-1}$ 
(Bennett et al. 2003) and {q$_0$}~=~1.}
Note that derived hydrogen densities vary by factor 1000 from one quasar to another. 
Yet there is a wide range of ionizing luminosities.
Meanwhile all derived distances of the \ion{Fe}{2} emitting region vary by only 
factor of 6.  Is it something common for all quasars, or are
these distances fortuitously similar to each other only in this small sample? 
Future reverberation mapping results for quasars over a range of luminosity 
would provide important comparison with the model predictions 
(c.f. Kaspi et al. 2000, hereafter K2000). We have one quasar 
in common with the K2000 sample, namely PG 1226. Its BLR distance based on
reverberation method 
is $\approx$ 7 times larger than that in our model. 
Note that the large distance for PG 1226 is not typical
for quasars in the K2000 sample, contrary to the distance derived 
in our model. On the other hand, the source for the discrepancy 
between the R found in our work for PG1226+023 and that obtained 
from reverberation mapping in K2000 may be quasar variability.

We have used our sample just for demonstration purposes and assumed   
a constant $v_{turb}$ = 5~km~s$^{-1}$ for all quasars.
By varying $v_{turb}$ we can achieve even better agreement between the observed
and predicted \ion{Fe}{2}(UV)/\ion{Mg}{2} ratios without any iron 
overabundance  (Verner et al. 2003a).

\section{Summary}
By applying our new 830-level model for Fe$^+$ ion, which is incorporated
into detailed photoionization calculations, we have probed the feasibility of using 
the strong 
\ion{Fe}{2} emission spectrum seen in quasars as a diagnostic tool for the physical 
conditions in BLRs. 
The \ion{Fe}{2} emission ratios show different trends as functions of the same 
parameters, 
hydrogen number density, and ionizing photon flux. We have found that the combination 
of the ratios is especially 
important in determining the hydrogen density, the ionizing flux and the radial 
distances to BLRs in quasars. 
We conclude that abundance is not the only factor that makes
\ion{Fe}{2} emission strong. Moreover,   
Fe abundance does not seem to be the dominant factor determining the strength of \ion{Fe}{2} emission. Our
modeling indicates that microturbulence, density, 
and radiation field have important roles and also lead to preferential 
strengthening of the \ion{Fe}{2} UV emission.

On the basis of the calculations presented here, we have three main
conclusions:

1. Our \ion{Fe}{2} large model atom used in photoionization calculations
predicts that large \ion{Fe}{2}(UV)/\ion{Mg}{2} ratios are not necessarily due 
to high iron abundance.
The \ion{Fe}{2}(UV)/\ion{Mg}{2} diagram (Figure 2) demonstrates that the
\ion{Fe}{2}(UV)/\ion{Mg}{2} ratios can have the same value over a wide range of 
physical conditions. 
This implies that the averaging of large numbers of quasars spectra may lead to  
properties of quasars that do not exist.

2. Since evolutionary models predict no iron overabundances for z $\leq$ 1,
observations to detect the optical iron band at these redshifts
will be beneficial to calibrate the model assuming a solar abundance for iron.
Such research must be accomplished with STIS and NICMOS aboard the HST to properly span the wavelength range where \ion{Fe}{2} 
emission is seen and  to obtain constraints on the most uncertain parameter, 
microturbulence.

3. Our \ion{Fe}{2} emission band analysis demonstrates that \ion{Fe}{2} has a great 
potential for becoming an important diagnostic tool of 
the physical conditions in BLR's of quasars. We suggest three 
\ion{Fe}{2} emission ratios, which can be used to obtain 
hydrogen density, flux of hydrogen-ionizing photons, microturbulent velocities,
and distances to BLR's.

In this paper we have shown that even in the limited space of 
parameters (ionizing flux and hydrogen density) exact fits 
to observed spectra exist.
The full investigation of our model in the whole parameter 
space including dust properties, Balmer continuum, and many 
others will be a subject of coming papers. Such investigation 
will provide all BLR physical parameters with the determined uncertainties. 



\acknowledgments

The research of EV has been supported through an NSF grant 
(NSF - 0206150) to CUA. SJ is supported by a grant from the 
Swedish National Space Board. We wish 
to acknowledge the use of the computational facilities of the 
Laboratory for Astronomy and Solar Physics (LASP) 
at NASA/Goddard Space Center. We give special thanks to Keith 
Feggans, Don Lindler, and Terry Beck for their computer 
services support. We are grateful to unknown referee for helpful comments.

\clearpage

\figcaption [f1.ps]{\ion{Fe}{2} pseudo-continuum predicted by 830-level
Fe$^+$ model for physical conditions in QSO BLRs.
The 2200$-$6000~{\AA} wavelength range is
 divided into small bins of 6~{\AA} (consistent with a $\sim 500$
km~s$^{-1}$ FWHM at 3600~{\AA})
and \ion{Fe}{2} flux calculated in each bin. The model at
$v_{turb}~=~0$~km~s$^{-1}$ predicts many strong
and narrow \ion{Fe}{2} features in the 1600$-$3500~{\AA} range (solid
line). The same model, but at $v_{turb}~=~5$~km~s$^{-1}$ is marked by dashed lines.}

\figcaption [f2.ps]{The \ion{Fe}{2}(UV)/\ion{Mg}{2} ratios predicted using
830-level
model for
Fe$^+$ in a QSO BLR as a function of the hydrogen density $n_H$
and flux of hydrogen-ionizing photons.
The chemical abundances are solar, the cloud column density is
$10^{24}$~cm$^{-3}$,
and  $v_{turb}~=~0,~5, $ and 10 ~km~s$^{-1}$.  }

\figcaption [f3.ps]{The predicted 
\ion{Fe}{2}(UV)/\ion{Fe}{2}(Opt1)
ratios calculated for the same parameters as in Figure 2.}

\figcaption [f4.ps]{The predicted
\ion{Fe}{2}(UV)/\ion{Fe}{2}(Opt2)
ratios calculated for same parameters as in Figure 2.}

\figcaption [f5.ps]{Contribution of 
\ion{Fe}{2} to \ion{Mg}{2} doublet, 
\ion{Fe}{2}(2720$-$2860{\AA})/\ion{Mg}{2},
$v_{turb} = 5$~km~s$^{-1}$.}

\newpage
\rotatebox{-90}{
\epsscale{0.9}
\plotone{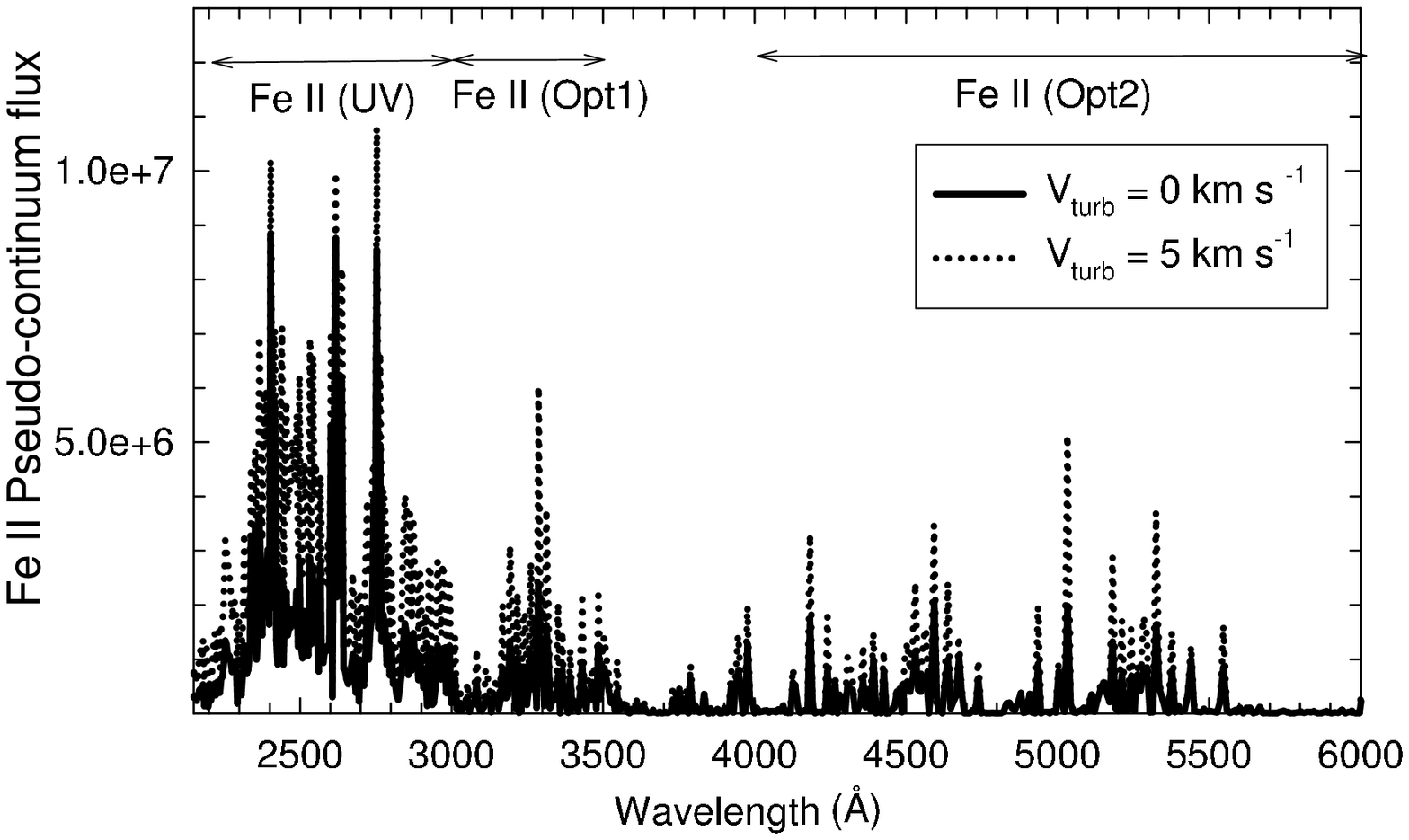}}
\plotone{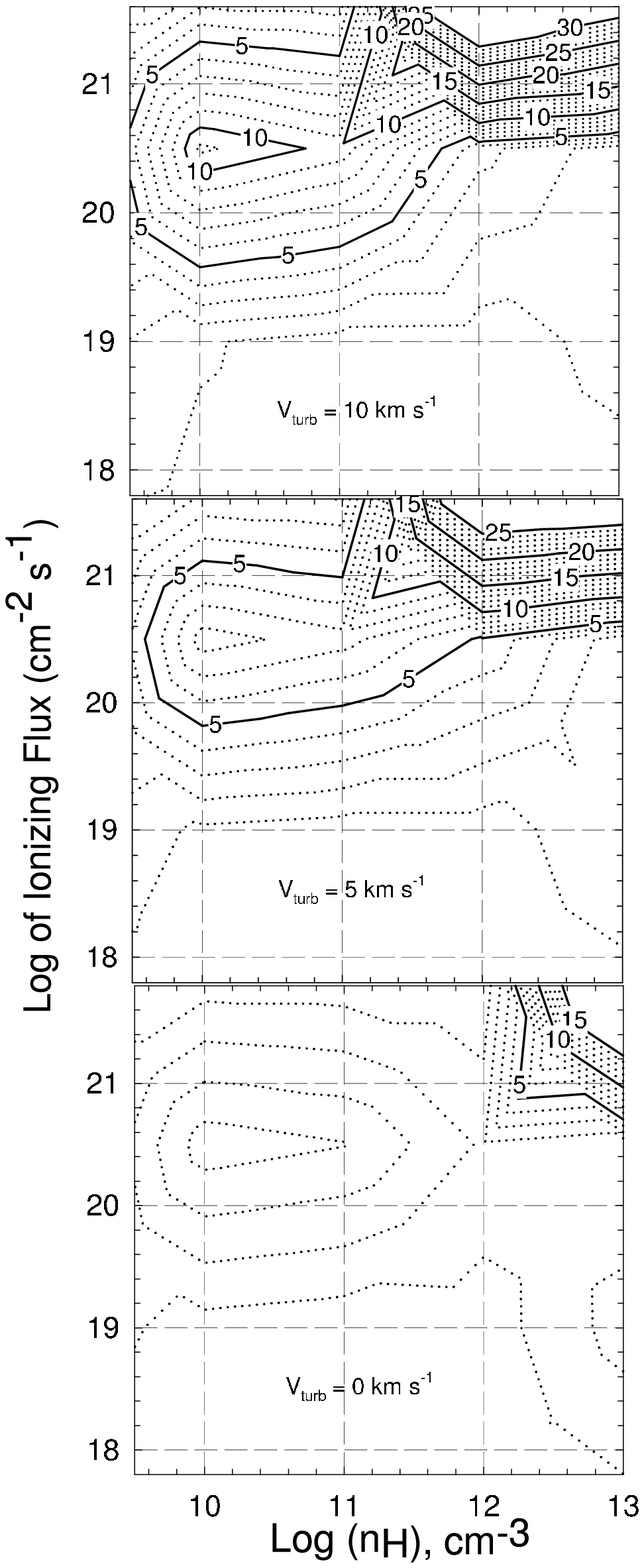}
\plotone{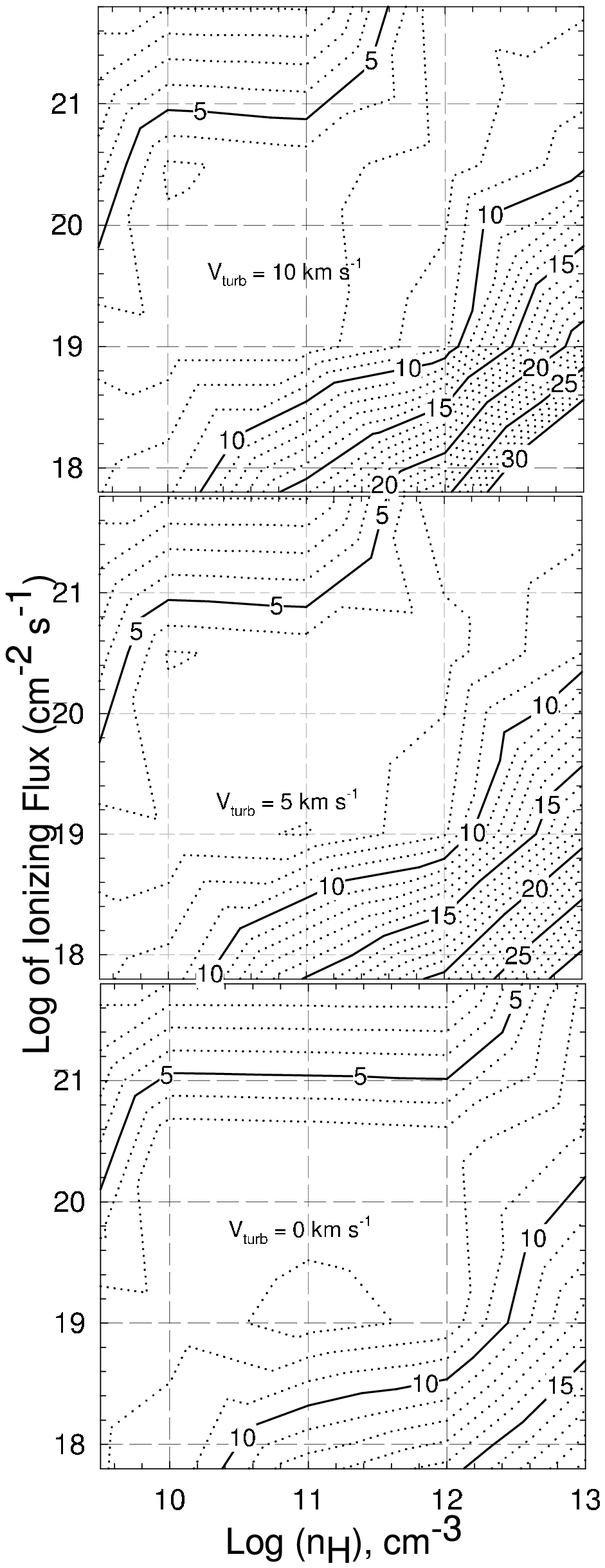}
\plotone{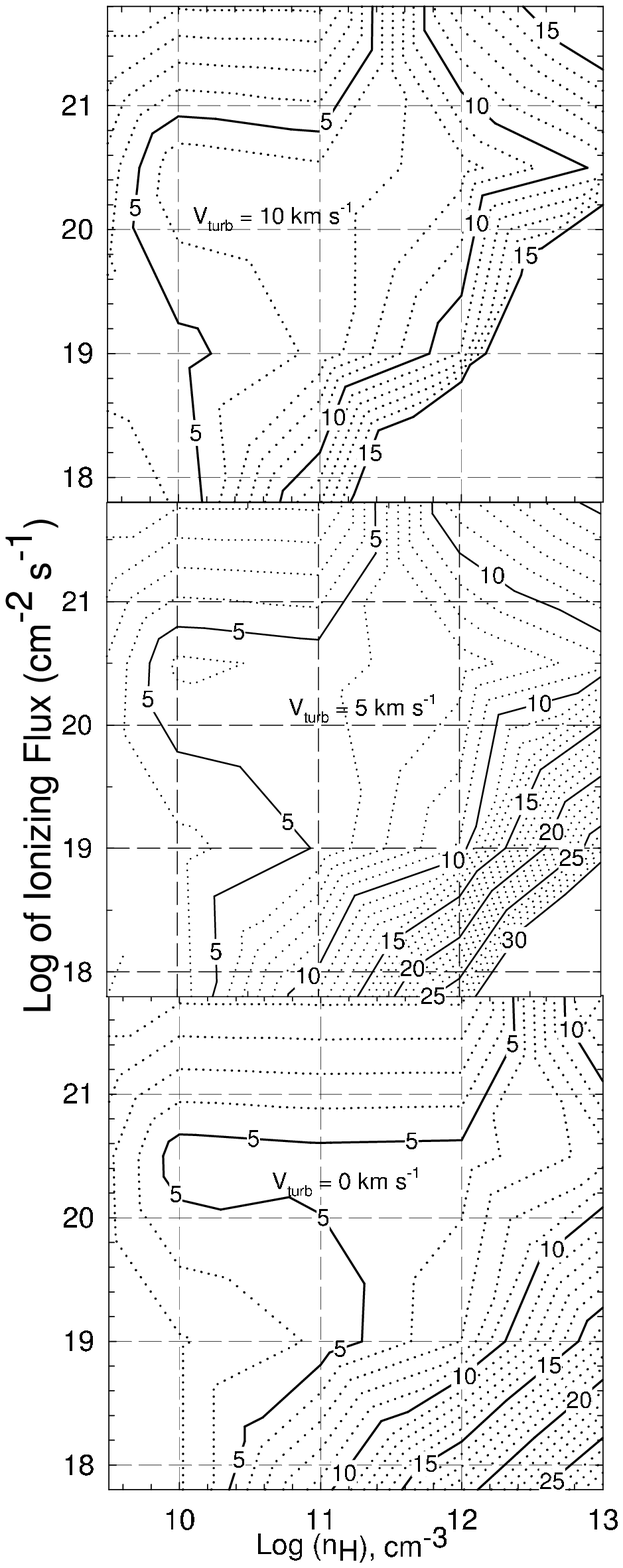}
\rotatebox{-90}{
\epsscale{0.9}
\plotone{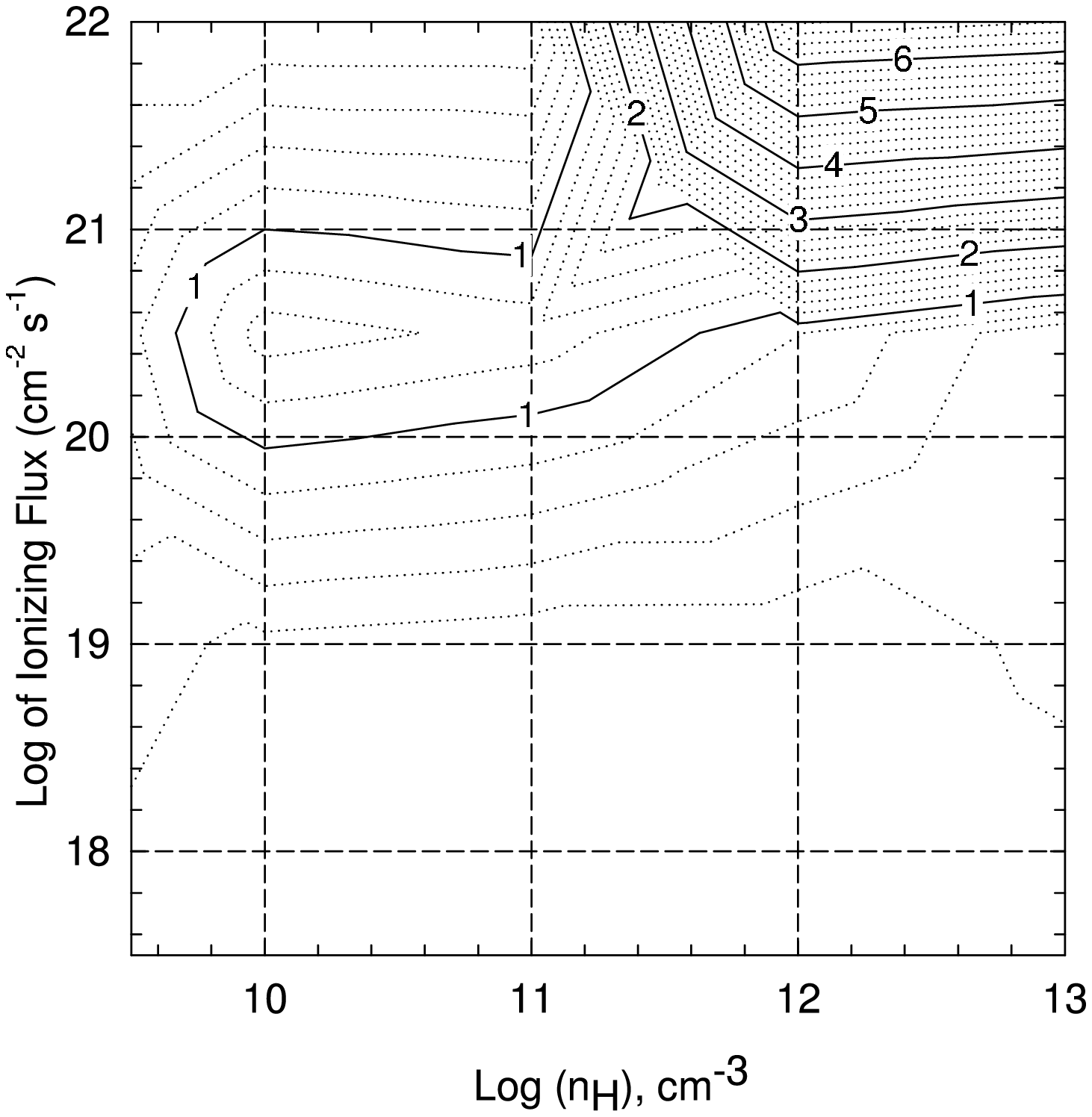}}






\clearpage


        
\begin{deluxetable}{cccccccc}
\rotate
\tablecaption{Predicted Hydrogen Densities, Ionizing Fluxes, \& Distances 
to BLRs.\label{tbl-2}}
\tablewidth{0pt}
\tablehead{
\colhead{}    &  \multicolumn{2}{c}{WNW85} &   \colhead{}   &
\multicolumn{2}{c}{This work} \\
\cline{1-4} \cline{5-8}\\
\colhead{Object} &
\colhead{$\mathrm{\frac{FeII(UV)}{MgII}}$}   &
\colhead{$\mathrm{\frac{FeII(UV)}{FeII(Opt1)}}$}   &
\colhead{$\mathrm{\frac{FeII(UV)}{FeII(Opt2)}}$}   &
\colhead{log n$_H$} &
\colhead{log $\Phi_{ion}$} &
\colhead{L\tablenotemark{b}} &
\colhead{R} \\

\colhead{} &
\colhead{} &
\colhead{} &
\colhead{} &
\colhead{cm$^{-3}$} &
\colhead{cm$^{-2}$~s$^{-1}$} &
\colhead{10$^{56}$~photons~s$^{-1}$} &
\colhead{10$^{17}$cm} \\
\cline{1-4} \cline{5-8}\\
\colhead{1} &
\colhead{2} &
\colhead{3} &
\colhead{4} &
\colhead{5} &
\colhead{6} &
\colhead{7} &
\colhead{8} \\
}
\startdata
0405$-$123 & 8 & 7.7 & 4 & 10 & 20.6 & 7.47 & 3.86\\
0738+313 & 5.4 & 5.0 & 4.8 & 10 & 21 & 5.08 & 2.01\\
0742+318\tablenotemark{a} & 6.1 & 13.2 & 8 & 13 & 20.5 & .57 & 1.20\\
1104+167 & 6.0 & 4.8 & 9.5 & 11.5 & 20.5 & 17.4 & 6.62\\
1226+023\tablenotemark{a} & 9.8 & 6.8 & 4.7 & 11.5 & 21.5 & 7.33 & 1.36\\
\enddata

\tablenotetext{a}{~Intensities have been corrected due to intrinsic reddening.}
\tablenotetext{b}{~Photon luminosity in 0.0136$-$10 keV range. See also text for details.}

\end{deluxetable}



\end{document}